\documentclass[12pt,fleqn]{article}
\usepackage[dvips]{graphics}
\title{Four-Spinor Reference Sheets}  
\date{ }
\author{{\it Richard Shurtleff~}\thanks{affiliation and mailing 
address: Department of Mathematics and Applied Sciences, 
Wentworth Institute of Technology, 550 Huntington Avenue, 
Boston, MA, USA, telephone number: (617) 989-4338, fax 
number: (617) 989-4591 , e-mail address: shurtleffr@wit.edu}} 
\begin{document}           
\maketitle               
			\begin{abstract}  %Abstract
Some facts about 4-spinors listed and discussed. None, well perhaps some, of the work is original. However, locating formulas in other places has proved a time-consuming process in which one must always worry that the formulas found in any given source assume the other metric ( I use $\{-1,-1,-1,+1\}$) or assume some other unexpected preconditions. Here I list some formulas valid in general representations first, then formulas using a chiral representation are displayed, and finally formulas in a special reference frame (the rest frame of the `current' $j$) in the chiral representation are listed. Some numerical and algebraic exercises are provided.

			\end{abstract}
		\section{General Representation}	% 1

	We can use any four complex numbers as the components of a 4-spinor in a given representation, $\psi$ = col$\{a+bi, c+di, e+fi, g+hi\}$, where `col' indicates a column matrix and the eight numbers $a$...$h$ are real. The 4-spinor generates four real-valued vectors: two light-like, one time-like and one space-like. These may be defined using the gamma matrices of the representation as follows:
		\begin{equation} 	% 1
j^\mu \equiv \overline{\psi }\gamma ^\mu \psi \,\,\,\,;\,\,\, a^\mu \equiv \overline{\psi }\gamma ^\mu \gamma ^5 \psi \,\,\,\,;\,\,\,\, r^\mu \equiv
\overline{\psi } \gamma ^\mu \left( \frac{1 + \gamma ^5 }2 \right) \psi \,\,\,\,;\,\,\,\,\,\, s^\mu \equiv \overline{\psi }\gamma ^\mu \left(
\frac{1-\gamma ^5} 2 \right) \psi ,
		\end{equation}	% 1
where $\overline{\psi } \equiv$ $\psi^{\dagger} \gamma^4$, $\mu$ is one of \{1,2,3,4\}, and $\gamma^5 \equiv$ $-i\gamma^1 \gamma^2 \gamma^3 \gamma^4$. Note that the vectors are representation independent; the substitution $\gamma^\mu \rightarrow$ $S^{-1}\gamma^\mu S$ and $\psi \rightarrow$ $S^{-1} \psi$ doesn't change the vectors.  By using a specific representation, perhaps the one displayed below in (3), one can show after some algebra that (i) $r$ and $s$ are light-like vectors and that (ii) $j$ is time-like and that (iii) $a$ is space-like. An exception occurs (iv) when $r$ or $s$ is zero; then $j$ and $a$ are light-like.

			\begin{figure}[h]	% 1
\vspace*{2in}
\hspace{1.25in}\includegraphics[0,144][288,0]{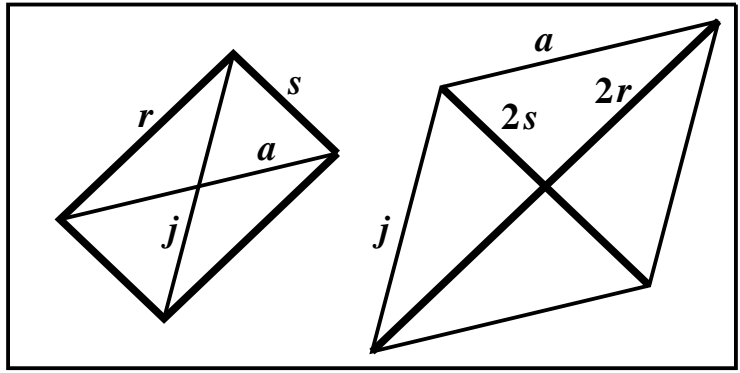}
\caption{The vectors make parallelograms. }
			\end{figure}

	Since the gammas in (1) are sandwiched between common factors of $\overline{\psi }$ and $\psi$, we see that the following are true:
		\begin{equation}	% 2
 j^\mu = r^\mu + s^\mu \,\,;\,\,\, a^\mu = r^\mu - s^\mu \,\,\,;\,\,\, 2 r^\mu = j^\mu + a^\mu \,\,\,;\,\,\, 2 s^\mu = j^\mu - a^\mu .
		\end{equation}	% 2
The vectors can be arranged in parallelograms, see Fig. 1.

	The scalar product of $j$ with itself, $j^2 \equiv$ $j^{\mu} j_{\mu}$, is the same as that for $a$, $a^{\mu} a_{\mu}$ = $-j^2$, except for the sign. The two vectors are `orthogonal', $j^{\mu} a_{\mu}$ = 0. We collect scalar products in Table 1.
\begin{table}[h] \parbox{2.5in}{\caption{Scalar products.}}\\
\begin{tabular}{r|cccc} \hline
Vector & $j$ & $a$ & $r$ & $s$ \\[0.5ex]
\hline
$j$ & $j^2$ & 0 & $j^2/2$ & $j^2/2$ \\
$a$ &  & $-j^2$ & $-j^2/2$ & $j^2/2$ \\
$r$ & & & 0 & $j^2/2$ \\
$s$ & & & & 0 
\end{tabular}
\end{table}
\vspace{.5in}

\pagebreak

\section{Chiral Representation [CR]}		% 2

	To get specific formulas for the vectors in terms of the components of the 4-spinor $\psi$ one must choose a representation for the gammas. I choose a chiral representation [CR]:
			\begin{equation} 	% 3
  \gamma^k = \pmatrix{ 0 & + \sigma^k e^{ i \delta } \cr - \sigma^k e^{ - i \delta } & 0 } ; \gamma^4  = \pmatrix{ 0 & - e^{ i \delta } \cr - e^{ - i \delta } & 0 }  ; \gamma^5 =  \pmatrix{ 1 & 0 \cr 0 & - 1} , \; \: [\mathrm{CR}]
 			\end{equation} 	% 3 
where $ \delta $ is an arbitrary phase angle, $k$ is any one of \{1,2,3\}, `1' is the unit 2x2 matrix, and the Pauli matrices are the 2x2 matrices
		\begin{equation}	% 4
 \sigma^1 =  \pmatrix{0&1 \cr 1&0}, \sigma^2 = \pmatrix{0&-i \cr i&0}, \sigma^3 = \pmatrix{1&0 \cr 0&-1}.  
		\end{equation}	% 4
One may check that the gammas (3) satisfy $\gamma^{\mu} \gamma^{\nu} + $ $\gamma^{\nu} \gamma^{\mu} $ = $2 g^{\mu \nu} \cdot 1$, where `1' is the unit 4x4 matrix and $ g^{\mu \nu}$ = diag$\{-1,-1,-1,+1\}$ is the 4x4 metric tensor.

	Write the 4-spinor $\psi$ as follows
		\begin{equation}	% 5
\psi =  \pmatrix{r \cos (\theta_R/2) \exp (-\frac{i\phi_R}{2}) \exp (i\frac{\alpha - \beta }{2}) \cr r \sin (\theta_R/2) \exp (+\frac{i\phi_R}{2}) \exp (i\frac{\alpha - \beta }{2}) \cr l \cos (\theta_L/2) \exp (-\frac{i\phi_L}{2}) \exp (i\frac{\alpha + \beta }{2}) \cr l \sin (\theta_L/2) \exp (+\frac{i\phi_L}{2}) \exp (i\frac{\alpha + \beta }{2}) } . \;\: \;\:  [\mathrm{CR}]
		\end{equation}	% 5
The given four complex numbers making up the components of $\psi$ determine the eight real numbers $r$, $\theta_R$, $\phi_R$, $l$, $\theta_L$, $\phi_L$, $\alpha$, and  $\beta$, within the usual additive $n\pi$'s. By (1), (3), and (5) one finds an expression for $j^2$:
		\begin{equation}	% 6
 j^2  = 2 r^2 l^2 ( 1 + \cos \theta_R \cos \theta_L + \cos \phi_R \cos \phi_L \sin \theta_R \sin \theta_L + \sin \phi_R \sin \phi_L \sin \theta_R \sin \theta_L ). 
		\end{equation}	% 6
\begin{flushright} [CR] \hspace*{0.3in}    \end{flushright}

By (1), with the parameters in (5) and the representation (3), one finds specific formulas for $r$ and $s$,
		\begin{equation}	% 7
 \{ r^1, r^2, r^3, r^4 \} = \{ r^2 \sin \theta_R \cos \phi_R , r^2 \sin \theta_R \sin \phi_R , r^2 \cos \theta_R, r^2 \}; \;\: \;\:  [\mathrm{CR}]
		\end{equation}	% 7
		\begin{equation}	% 8
 \{ s^1, s^2, s^3, s^4 \} = \{ - l^2 \sin \theta_L \cos \phi_L , - l^2 \sin \theta_L \sin \phi_L , - l^2 \cos \theta_L, l^2 \}. \;\: \;\:  [\mathrm{CR}]
		\end{equation}	% 8
Clearly the angles $\theta$ and $\phi$ are polar and azimuthal angles of the spatial directions of $r$ and $s$. Specific formulas for $j$ and $a$ follow immediately from (2), (7), and (8).

	With the chiral representation the 4-spinor splits into two 2-spinors, $\psi$ = col$\{ \rho, \lambda \}$, where `col' means column matrix. The 2-spinor $\rho$ is right-handed and the other, $\lambda$, is left-handed, referring to their Lorentz transformation properties. By (5), (7), and (8) one sees that the right 2-spinor $\rho$ determines $r$ and the left 2-spinor $\lambda$ determines $s$. The 2x2 rotation matrix $R(\kappa, \mathbf{\hat{n}})$ for a rotation through an angle $\kappa$ about the direction $ \mathbf{\hat{n}}$ is the same for both right and left 2-spinors, $R(\kappa, \mathbf{\hat{n}})$ = $\exp (-i \mathbf{\hat{n}}_k \sigma^{\mathit{k}} \kappa/\mathrm{2})$. The 2x2 boost matrix $B(u, \mathbf{\hat{n}})$ for a boost of speed $ \tanh u$ in the direction $ \mathbf{\hat{n}}$ differs for right and left 2-spinors: $B_R (u, \mathbf{\hat{n}})$ = $\exp (+ \mathbf{ \hat{n} }_k \sigma^{\mathit{k}} \mathit{u}/\mathrm{2})$ and $B_L (u, \mathbf{\hat{n}})$ = $\exp (- \mathbf{ \hat{n} }_k \sigma^{\mathit{k}} \mathit{u}/\mathrm{2})$.

	A rotation through an angle $\kappa$ about the direction $ \mathbf{\hat{n}}$ changes the 4-spinor $\psi$: $\psi \rightarrow$ $ [\cos (\kappa/2) \cdot 1 - $ $i \sin (\kappa/2) n_k \gamma^5 \gamma^4 \gamma^k ] \psi$, where `1' is the unit 4x4 matrix. The rotation through $\kappa$ about $ \mathbf{\hat{n}} $ = \{0,0,1\} changes $\{j^1, j^2 \}$ to $ \{ \cos \kappa j^1 - \sin \kappa j^2 , \sin \kappa j^1 + \cos \kappa j^2 \}$, leaving $j^3$ and $j^4$ unchanged. 

	A boost of speed $ \tanh u$ in the direction $ \mathbf{\hat{n}}$ changes the 4-spinor $\psi$: $\psi \rightarrow$ $ [\cosh (u/2) \cdot 1 + $ $ \sinh (u/2) n_k \gamma^4 \gamma^k ] \psi$, where `1' is the unit 4x4 matrix. The boost of speed $ \tanh u$ in the direction $ \mathbf{\hat{n}}$ = \{0,0,1\} changes $\{j^3, j^4 \}$ to $ \{ \cosh u j^3 + \sinh u j^4 , \sinh u j^3 + \cosh u j^4 \}$, leaving $j^1$ and $j^2$ unchanged. 

\section{$j$-time frame}

	By applying the appropriate boost (3 parameters: $u$, $\hat{n}^1$, $\hat{n}^2$ which determines $\hat{n}^3$) we get a new $j$ which has no spatial components; the new $j$ is in its proper frame. Call this the `$j$-time frame.' In this frame the spinor has equal right and left 2-spinors within a phase, $\rho$ = $e^{-i \beta} \lambda$, and the light-like vectors $r$ and $s$ point in opposite directions. The transformed 4-spinor may be written in the form
		\begin{equation}	% 9
 \psi = \sqrt{{\frac{j}{2}}} \pmatrix{\cos (\theta/2) \exp (-\frac{i\phi}{2}) \exp (i\frac{\alpha - \beta }{2}) \cr \sin (\theta/2) \exp (+\frac{i\phi}{2}) \exp (i\frac{\alpha - \beta }{2}) \cr \cos (\theta/2) \exp (-\frac{i\phi}{2}) \exp (i\frac{\alpha + \beta }{2}) \cr \sin (\theta/2) \exp (+\frac{i\phi}{2}) \exp (i\frac{\alpha + \beta }{2}) } , \;\: \;\:  [\mathrm{CR}]
		\end{equation}	% 9
(i) [$\{ \theta, \phi \}$] where $\{ \theta, \phi \}$ are the \{ polar, azimuthal \} angles indicating the direction of $\mathbf{r}$ and $\mathbf{a}$ which is opposite to the direction of $\mathbf{s}$. The overall phase is $\alpha /2$ and the phase shift from the right 2-spinor to the left 2-spinor is $\beta$. The four angles $\{ \theta, \phi, \alpha, \beta \}$, the magnitude of $j$, and the three parameters $u$, $\hat{n}^1$, $\hat{n}^2$ of the boost amount to eight real numbers which is the same number needed to specify the four complex numbers making up a 4-spinor in a given representation. Thus we still have a general form for the 4-spinor.

	(ii) [$\alpha$] Rotating $\psi$ in the $j$-time frame, (9), leaves $j$ alone and changes the values of $\{ \theta, \phi, \alpha\}$. If the rotation axis is in the direction of $\mathbf{a}$, $\hat{n}^k$ = $a^k/ \sqrt{j^2 + (a^4)^2}$ with $a^4$ = 0 in this frame, then the effect on $\alpha$ is especially simple: $\alpha$ changes by the negative of the rotation angle $\kappa$, $\alpha \rightarrow$ $\alpha - \kappa$. Rotating by $\kappa$ = $\alpha$ about $\mathbf{a}$ brings $\alpha$ to zero, $\alpha \rightarrow$ 0. Therefore we may interpret $\alpha$, twice the overall phase of $\psi$ in this frame, as a rotation angle.

	The way this works can be seen as follows. When the direction $\mathbf{a}$ is along \{1,0,0\}, the angles $\theta$ and $\phi$ in (9) are $\theta$ = $\pi /2$ and $\phi$ = 0 or $\pi$. For $\phi$ = 0 the right and left 2-spinors are given by $\rho$ = $\lambda$ =  $\exp (i \alpha / 2 )$ col\{1,1\} if we take $\beta$ = 0 and $j$ = 4. As noted above, the effect of a rotation is to multiply both $\rho$ and $\lambda$ by the same 2x2 matrix $R(\kappa, \mathbf{\hat{n}})$. The rotation matrix $\exp (-i \sigma^{\mathrm{1}} \kappa/\mathrm{2})$ for $\mathbf{\hat{n}}$ = \{1,0,0\} is a linear combination of the Pauli matrix $\sigma^1$ and the unit 2x2 matrix. But the 2-spinors are eigenspinors of $\sigma^1$ and the unit 2x2 matrix with eigenvalue 1, so the effect of the rotation matrix $\exp (-i \sigma^{\mathrm{1}} \kappa/\mathrm{2})$ is to change the phase of $\rho$ and $\lambda$ by $-\kappa /2$. In short, the two 2-spinors are eigenspinors of the rotation matrix with the same eigenvalue which is the common phase factor $ \exp (-i \kappa /2)$. 

For $\phi$ = $\pi$, the 2-spinor $\rho$ = $\lambda$ =  $\exp (i \alpha / 2 )$ col\{-1,1\} is an eigenspinor of $\sigma^1$ with eigenvalue $-1$, so the common phase factor is $ \exp (+i \kappa /2)$. In Table 2, we collect the change in angles $\{ \theta, \phi, \alpha\}$ due to rotations of angle $\kappa$ about the coordinate axes. 

	(iii) [$\beta$] The phase $\beta$ is changed, $\beta \rightarrow$ $\beta \pm \kappa$ sign depending on eigenvalue, when the right-handed 2-spinor $\rho$ is rotated by $\kappa$ and $\lambda$ is rotated through $-\kappa$, both rotations taking place about $\mathbf{a}$. In this case none of the angles $\{ \theta, \phi, \alpha\}$ changes and the magnitude of $j$ doesn't change.

	(iv) [$j$] An operation that changes only the magnitude of $j$ while leaving $\{ \theta, \phi, \alpha, \beta\}$ alone can be found. If the right 2-spinor $\rho$ is boosted along the direction of $\mathbf{a}$ by $ \tanh u$ and $\lambda$ is boosted by the same speed but in the opposite direction $-\mathbf{a}$, then the magnitude of $j$ changes, $j \rightarrow$ $[ \cosh u - \sinh u ] j$. 

	Thus the 4-spinor parameters $\{ \theta, \phi, \alpha\}$ can each be changed by a suitable rotation applied to $\psi$, $\beta$ alone can be changed by applying a counter-clockwise rotation to the right-handed 2-spinor $\rho$ and the equal clockwise rotation to $\lambda$, and the magnitude of $j$ alone can be changed by boosting $\rho$ forward and boosting $\lambda$ backward.

\begin{table}[h] {\caption{Changes $\{ \Delta \theta, \Delta \phi, \Delta \alpha\}$ due to a rotation of angle $\kappa$ about each coordinate axis. Values of $\{ \theta, \phi, \alpha\}$ are provided that give the components of the eigenspinors. The $x^1$ and $x^2$ eigenspinors are not normalized.} }
\vspace{0.2in}
\begin{tabular}{r|cccc} \hline
Eigenspinor $\rightarrow$ & $x^1$ & $x^1$ & $x^2$ & $x^2$ \\[0.5ex]
Components $\rightarrow$ & col$\{-1,1\}$ & col$\{1,1\}$ & col$\{i,1\}$ & col$\{-i,1\}$ \\[0.5ex]
$\{ \theta, \phi, \alpha\}$ $\rightarrow$ & $\{ \frac{\pi}{2}, \pi, -\pi \}$ & $\{ \frac{\pi}{2}, 0,0 \}$ & $\{ \frac{\pi}{2}, -\frac{\pi}{2}, \frac{\pi}{2} \}$ & $\{ \frac{\pi}{2}, \frac{\pi}{2}, -\frac{\pi}{2} \}$  \\[0.5ex]
Rotation Axis $\downarrow $ & & & & \\[0.5ex]
\hline $x^1$-axis & $\{0,0,+\kappa\}$ & $\{0,0,-\kappa\}$ & $\{+\kappa,0,0 \}$ & $\{-\kappa,0,0 \}$ \\
$x^2$-axis & $\{-\kappa,0,0 \}$ & $\{+\kappa,0,0 \}$ & $\{0,0,+\kappa\}$ & $\{0,0,-\kappa\}$ \\
$x^3$-axis & $\{0,+\kappa, 0\}$ & $\{0,+\kappa, 0\}$ & $\{0,+\kappa, 0\}$ & $\{0,+\kappa, 0\}$ 
 
\end{tabular}
\end{table}

\begin{table}[h] {\caption{A continuation of Table 2} }
\vspace{0.2in}
\begin{tabular}{r|cc} \hline
Eigenspinor $\rightarrow$ &  $x^3$ & $x^3$ \\[0.5ex]
Components $\rightarrow$ &  col$\{0,1\}$ & col$\{1,0\}$\\[0.5ex]
$\{ \theta, \phi, \alpha\}$ $\rightarrow$ & $\{ \pi, \phi_{0}, -\phi_{0} \}$ & $\{ 0, \phi_{0}, \phi_{0} \}$ \\[0.5ex]
Rotation Axis $\downarrow $ & & \\[0.5ex]
\hline $x^1$-axis & $\{-\kappa,-\phi_{0} +\frac{\pi}{2},+\phi_{0} - \frac{\pi}{2} \}$ & $\{+\kappa,-\phi_{0} - \frac{\pi}{2},-\phi_{0} - \frac{\pi}{2} \}$ \\
$x^2$-axis & $\{-\kappa,-\phi_{0} +\pi,+\phi_{0} - \pi \}$ & $\{+\kappa,-\phi_{0},-\phi_{0} \}$ \\
$x^3$-axis & $\{0,0,+\kappa\}$ & $\{0,0,-\kappa\}$
\end{tabular}
\end{table}

\pagebreak

\clearpage

\appendix

\section{Problems} % 4

1. Find $j$, $a$, $r$, and $s$ when 

(i) the 4-spinor $\psi$ has four equal real-valued components: $A$ = $a$ = $c$ = $e$ = $g$ and 0 = $b$ 

= $d$ = $f$ = $h$; 

(ii) as in (i) but with $c$ negative: $A$ = $a$ = $-c$ = $e$ = $g$ and 0 = $b$ = $d$ = $f$ = $h$; 

(iii) try $A$ = $a$ = $d$ = $e$ = $f$, 0 = $b$ = $c$, and $2A$ = $g$.\smallskip

\noindent2. Use the gammas (3) to find $j$ as a function of $a$ ... $h$.\smallskip

\noindent3. Show that $\gamma^1 \cdot \gamma^2 + \gamma^2 \cdot \gamma^1$ = 0 and that $\gamma^2 \cdot \gamma^2 + \gamma^2 \cdot \gamma^2$ = $-2 \cdot 1$, where `1' is the unit 4x4 

matrix.\smallskip

\noindent4. By definition, $\exp [-i  \sigma^1 \kappa/2] \equiv$ $ \Sigma ( - i \sigma^1 \kappa/2 )^n/n!$. 

(i) Calculate $(\sigma^1)^2$ = $\sigma^1 \cdot \sigma^1$.

(ii) Show $\exp [-i  \sigma^1 \kappa/2]$ = $\cos (\kappa/2) \cdot 1$ - $i \sin(\kappa/2) \sigma^1$, where `1' is the unit 2x2 matrix.\smallskip

\noindent5. Find $r$, $\theta_R$, $\phi_R$, $\alpha$, $\beta$, $l$, $\theta_L$, and $\phi_L$ for the 4-spinor of problem 1(iii).\smallskip

\noindent6. The parity operator $P$ has the following effect on a 4-spinor in the chiral representation: 

$P \pmatrix{\rho \cr \lambda}$ = $\pmatrix{- \lambda \cr - \rho}$, where $\rho$ and $\lambda$ are the right- and left-handed 2-spinors. The charge

conjugation operator $C$ has the following effect:  $C \psi$ = $i \gamma^2 \psi$. 

Apply $P$, $C$ and $CP$ to the 4-spinor of problem 1(iii) and find the $j$'s and $a$'s.\smallskip

\noindent7. (i) Find a 64 component quantity $\Gamma^{\mu}_{\nu \, \tau}$ so that $j^\mu$ = $ -\Gamma^{\mu}_{\nu \, \tau} r^\nu s^\tau$ and $\Gamma^{\mu}_{\nu \, \tau}$ = $ - \Gamma^{\mu}_{\tau \, \nu}$. 

(ii) Show that 0 = $r^\mu$ + $s^\mu$ + $\Gamma^{\mu}_{\nu \, \tau} r^\nu s^\tau$. Interpret that equation using parallel transfer 

and the parallelograms of Figure 1.

%
%%%%%%%%%%   Reference   %%%%%%%%%%
%
%\pagebreak

\end{document}